\documentclass[,final]{aipproc}

\layoutstyle{8x11single}

\usepackage{epsfig}
\usepackage[dvips]{color}
\usepackage{amsmath}
\usepackage{amssymb}

\def\beq{\begin{equation}}
\def\eeq{\end{equation}}

\def\bea{\arraycolsep .1em \begin{eqnarray}}
\def\eea{\end{eqnarray}}
\def\Tr{{\rm Tr}}

\def\la{\lambda}
\def\lab{\lambda_{\rm bound}}

\def\s0#1#2{\mbox{\small{$ \frac{#1}{#2} $}}}
\def\0#1#2{\frac{#1}{#2}}

\def\grgl{\:\hbox to -0.2pt{\lower2.5pt\hbox{$\sim$}\hss}{\raise3pt\hbox{$>$}}\:}
\def\klgl{\:\hbox to -0.2pt{\lower2.5pt\hbox{$\sim$}\hss}{\raise3pt\hbox{$<$}}\:}

\begin{document}

\title{Fixed points of quantum gravity in higher dimensions}

\classification{04.50.+h, 04.60.-m, 11.15.Tk}
\keywords      {Quantum gravity, Extra dimensions}

\author{P.~Fischer}{
  address={Institut f\"u{}r Theoretische Physik E, RWTH Aachen, D-52056 Aachen}}

\author{D.~F.~Litim}{
  address={School of Physics and Astronomy, University of Southampton,
  Southampton SO17 1BJ, U.K.\\ and Theory Group, CERN, CH-1211 Geneva 23}}

\begin{abstract}
  We study quantum gravity in more than four dimensions by means of an exact
  functional flow.  A non-trivial ultraviolet fixed point is found in the
  Einstein--Hilbert theory. It is shown that our results for the fixed point
  and universal scaling exponents are stable.  If the fixed point persists in
  extended truncations, quantum gravity in the metric field is asymptotically
  safe.  We indicate physical consequences of this scenario in
  phenomenological models with low-scale quantum gravity and large extra
  dimensions.
\end{abstract}

\begin{flushright}
SHEP-06-14, CERN-PH-TH-2006/066, PITHA 06/06
\end{flushright}

\maketitle

%%%%%%%%%%%%%%%%%%%%%%%%%%%%%%%%%%%%%%%%%%%%
%% MAINMATTER
%%%%%%%%%%%%%%%%%%%%%%%%%%%%%%%%%%%%%%%%%%%%

%**************************************************************************
\subsection{Introduction}  \label{Intro}
%**************************************************************************
In this contribution, we discuss new ultraviolet fixed points for quantum
gravity in higher dimensions \cite{Fischer:2006fz}.

Gravity in more than four space-time dimensions has received considerable
interest in recent years.  The possibility that the fundamental Planck mass --
within a higher dimensional setting -- may be as low as the electroweak scale
\cite{Arkani-Hamed:1998rs,Antoniadis:1990ew,Randall:1999ee} has stimulated
extensive model building and numerous investigations aiming at signatures of
extra spatial dimensions ranging from particle collider experiments to
cosmological and astrophysical settings.  Central to these scenarios is that
gravity lives in higher dimensions, while standard model particles are often
confined to the four dimensional brane.  In part, these models are motivated
by string theory, where additional spatial dimensions arise naturally. Then
string theory would, at least in principle, provide for a short distance
definition of these theories which presently have to be considered as
effective rather than fundamental ones.  In the absence of an explicit
ultraviolet completion, gravitational interactions at high energies including
low scale gravity can be studied with effective field theory or semi-classical
methods, as long as quantum gravitational effects are absent, or suppressed by
some ultraviolet cutoff of the order of the fundamental Planck mass, $e.g.$
\cite{Giudice:1998ck}.

One may then wonder whether a quantum theory of gravity in the metric degrees
of freedom can exist in four and more dimensions as a cutoff-independent,
well-defined and non-trivial local theory down to arbitrarily small distances.
It is generally believed that the above requirements imply the existence of a
non-trivial ultraviolet fixed point under the renormalisation group, governing
the short-distance physics. The corresponding fixed point action then
provides a microscopic starting point to access low energy phenomena of
quantum gravity.  This ultraviolet completion 
does apply for quantum
gravity in the vicinity of two dimensions, where an ultraviolet fixed point
has been identified with $\epsilon$-expansion techniques
\cite{Weinberg,Gastmans:1977ad,Aida:1996zn}.  In the last couple of years, a
lot of efforts have been put forward to access the four-dimensional case, and
a number of independent studies have detected an ultraviolet fixed point using
functional and renormalisation group methods in the continuum
\cite{Reuter:1996cp,Souma:1999at,Lauscher:2001ya,Lauscher:2002mb,Reuter:2001ag,Percacci:2002ie,Litim:2003vp,Bonanno:2004sy,Litim2005,Forgacs:2002hz}
and Monte Carlo simulations on the lattice
\cite{Hamber:1999nu,Ambjorn:2004qm}.

Continuity in the dimension suggests that a non-trivial fixed point -- if it
exists in four dimensions and below -- should persist at least in the vicinity
and above four dimensions.  Furthermore, the critical dimension of quantum
gravity -- the dimension where the gravitational coupling has vanishing
canonical mass dimension -- is two.  For any dimension above the critical one,
the mass dimension of the gravitational coupling is negative.  Hence, from a
renormalisation group point of view, four dimensions are not special. 
More generally, one expects that the local structure of quantum fluctuations,
and hence local renormalisation group properties of quantum theories of
gravity, are qualitatively similar for all dimensions above the critical one,
modulo topological effects in specific dimensions.

In this talk, we summarise our results for ultraviolet fixed points of gravity
in more than four dimensions, also extending the results given previously in
\cite{Litim:2003vp,Fischer:2006fz,Litim2005}.  For technical details, see
\cite{FischerLang,FischerGauge}.

%**************************************************************************
\subsection{Flows of quantum gravity}
%**************************************************************************

We perform a fixed point search for quantum gravity in more than four
dimensions \cite{Litim:2003vp} (see also \cite{Reuter:2001ag}). An ultraviolet
fixed point, if it exists, should already be visible in the purely
gravitational sector, to which we confine ourselves. Matter degrees of freedom
and gauge interactions can equally be taken into account.  We employ a
functional renormalisation group based on a cutoff effective action $\Gamma_k$
for the metric field
\cite{Reuter:1996cp,Souma:1999at,Lauscher:2001ya,Lauscher:2002mb,Reuter:2001ag,Percacci:2002ie,Litim:2003vp,Bonanno:2004sy,Litim2005,Branchina:2003ek},
see \cite{ERG-Reviews} and \cite{Litim:1998nf} for reviews in scalar and gauge
theories.  The functional $\Gamma_k$ comprises momentum fluctuations down to
the momentum scale $k$, interpolating between $\Gamma_\Lambda$ at some
reference scale $k=\Lambda$ and the full quantum effective action at $k\to 0$.
The variation of the effective action with the cutoff scale $(t=\ln k)$ is
given by an exact functional flow
\begin{equation}\label{ERG}
\partial_t\Gamma_k=
\s012\Tr\frac{1}{\Gamma_k^{(2)}+R_k}\partial_t R_k\,.
\end{equation}
The trace is a sum over fields and a momentum integration, and $R_k$ is a
momentum cutoff for the propagating fields.  The flow relates the change in
$\Gamma_k$ with a loop integral over the full propagator.
By construction, the flow \eqref{ERG} is well-defined (finite, no poles), and,
together with the boundary condition $\Gamma_\Lambda$, defines the theory.  In
renormalisable theories, the cutoff $\Lambda$ can be removed,
$\Lambda\to\infty$, and $\Gamma_\Lambda\to \Gamma_*$ remains well-defined for
arbitrarily short distances.  In perturbatively renormalisable theories,
$\Gamma_*$ is given by the classical action, $e.g.$ in QCD.  In perturbatively
non-renormalisable theories, proving the existence (or non-existence) of a
short distance limit $\Gamma_*$ is more difficult.  A fixed point action
qualifies as a fundamental theory if it is connected with the correct
long-distance behaviour by a well-defined renormalisation group trajectory
$\Gamma_k$.

To solve the flow \eqref{ERG}, we restrict $\Gamma_k$ to a finite set of
operators, which can systematically be extended. Highest reliability and best
convergence behaviour is achieved through an optimisation of the momentum
cutoff
\cite{Litim:2000ci,Litim:2002cf,Litim:2005us,FischerLang,Pawlowski:2003hq}.
We employ the Einstein-Hilbert truncation where the effective action, apart
from a classical gauge fixing and the ghost term, is given as
\begin{equation}\label{EHk}
\Gamma_k=
\0{1}{16\pi G_k}\int d^Dx \sqrt{g}\left[-R(g)+2\bar\lambda_k\right]\,.
\end{equation}
In \eqref{EHk}, $g$ denotes the determinant of the metric field $g_{\mu\nu}$,
$R(g)$ the Ricci scalar, 
$G_k$ the gravitational coupling constant, and $\bar\lambda_k$ the 
cosmological constant.  
For $k$-independent $G_k$ and $\bar\lambda_k$,
\eqref{EHk} reduces to the standard Einstein--Hilbert theory. 
The dimensionless renormalised gravitational and cosmological constants are
\begin{equation}\label{glk}
g=k^{D-2}\, G_k\, \equiv k^{D-2}\, Z^{-1}_{N,k}\ \bar G\, , \qquad
\lambda=\,k^{-2}\, \bar\lambda_k\ ,
\end{equation}
where $\bar G$ and $\bar\lambda_{\Lambda}$ denote the couplings at some
reference scale $k=\Lambda$, and $Z_{N,k}$ the wave function renormalisation
factor for the newtonian coupling.  Their flows are given by
\begin{equation}\label{dglk}
k \partial_k g\,       \equiv  \beta_g = \big[ D-2 + \eta_N  \big] g \, , \qquad
k \partial_k \lambda\, \equiv   \beta_\lambda
\end{equation}
with $\eta_N(\lambda,g)=-k \partial_k \ln Z_{N,k}$ the anomalous dimension of
the graviton.  Fixed points correspond to the simultaneous vanishing of
\eqref{dglk} at values $(\la_*, g_*)$ of the couplings.  Explicit expressions
for \eqref{dglk} and $\eta$ follow from \eqref{ERG} by projecting onto the
operators in \eqref{EHk}, using background field methods.  We employ momentum
cutoffs with tensorial structures introduced in \cite{Reuter:1996cp} (Feynman
gauge) and in \cite{Lauscher:2001ya} (harmonic background field gauge), and
optimised scalar cutoff function (see below).  For explicit analytical flow
equations, see \cite{Litim:2003vp,Litim2005}.  The ghost wave function
renormalisation is set to unity. We neglect the running of the gauge
fixing parameter $\alpha$ and study the $\alpha$-dependence of the results.
It is known that $\alpha=0$ is a non-perturbative fixed point of the flow
\eqref{ERG}, independently of the truncation \cite{Litim:1998qi}.
Diffeomorphism invariance is controlled by modified Ward identities
\cite{Reuter:1996cp}, similar to those employed for non-abelian gauge theories
\cite{Freire:2000bq}.

Three comments are in order. Firstly, the cosmological constant $\lambda$
obeys $\lambda<\lambda_{\rm bound}$, where $2\lambda_{\rm
  bound}=\min_{q^2/k^2}[(q^2+R_k(q^2))/k^2]$ depends on the momentum cutoff
$R_k(q^2)$ and $q^2\ge 0$ denotes (minus) covariant momentum squared. For
gauge fixing parameters $\alpha>1$, we have $\lambda_{\rm bound}\to \alpha
\,\lambda_{\rm bound}$, see \cite{Litim:2003vp,Litim2005}.  Elsewise the flow
\eqref{ERG}, \eqref{dglk} could develop a pole at $\lambda=\lambda_{\rm
  bound}$.  The property $\lambda<\lambda_{\rm bound}$ is realised in any
theory where $\Gamma_k^{(2)}$ develops negative eigenmodes, and simply states
that the inverse cutoff propagator $\Gamma_k^{(2)} +R_k$ stays positive
(semi-) definite \cite{Litim:2006nn}.  Secondly, we detail the momentum
cutoffs used in the numerical analysis. We introduce $R_k(q^2)=q^2\, r(y)$,
where $y=q^2/k^2$.  Within a few constraints these regulators can be chosen
freely \cite{Litim:2000ci}.  Presently, we use $r_{\rm mexp}=b/((b+1)^{y}-1)$,
$r_{\rm exp}=1/(\exp c y^b -1)$, $r_{\rm mod}=1/(\exp[c(y+(b-1)y^b)/b] -1)$,
with $c=\ln 2$, $r_{\rm pow}=y^{-b}$ and $r_{\rm opt}=b(1/y-1)\theta(1-y)$,
with $b>0$ for $r_{\rm mexp}$ and $r_{\rm opt}$ and $b\geq 1$ for the others.
These cutoffs include the sharp cutoff for $(b\to\infty)$ and asymptotically
smooth Callan-Szymanzik type cutoffs $R_k\sim k^2$ as limiting cases.  The
larger the parameter $b$, for each class of cutoffs, the `sharper' the
corresponding momentum cutoff, i.e. the narrower the window of momentum modes
contributing at each infinitesimal integration step in the renormalisation
flow.  Thirdly, we note that physical quantities, derived from the full,
integrated flow \eqref{ERG}, are independent of the cutoff $R_k$.
Truncations, however, may induce artificial cutoff dependences. It is vital to
establish which cutoffs lead to the most reliable results within a given
expansion. We employ a criterion based on the stability of the flow
\eqref{ERG}, which provides optimised choices for $R_k$ (or the parameter
$b$).  In the present case, stability is controlled by the parameter
$1-\lambda_k/\lambda_{\rm bound}$, which reads $1-\lambda_*/\lambda_{\rm
  bound}$ on a fixed point (see \cite{FischerLang} for details).

%*************************************************************************
\subsection{Results}  \label{results}
%*************************************************************************

Next we summarise our results for non-trivial ultraviolet fixed points
    $(g_*,\lambda_*)\neq(0,0)$ of \eqref{dglk}, the related universal scaling
    exponents, trajectories connecting the fixed point with the perturbative
    infrared domain, the graviton anomalous dimension, cutoff independence,
    gauge-fixing independence
    and the stability of the underlying expansion. We restrict ourselves to
    $D=4+n$ dimensions, with $n=0,...,7$.

%*************************************************************************
% Figure 1 
%*************************************************************************
\begin{figure}
\unitlength0.001\textwidth
\begin{picture}(1000,340)
\put(0,10){\epsfig{file=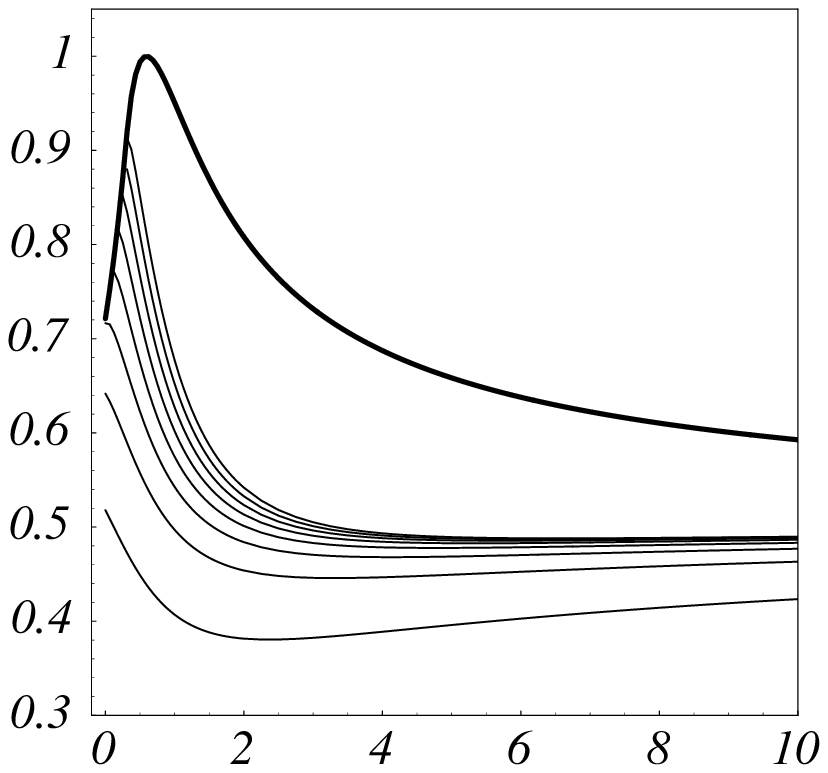,width=0.33\textwidth}}
\put(330,10){\epsfig{file=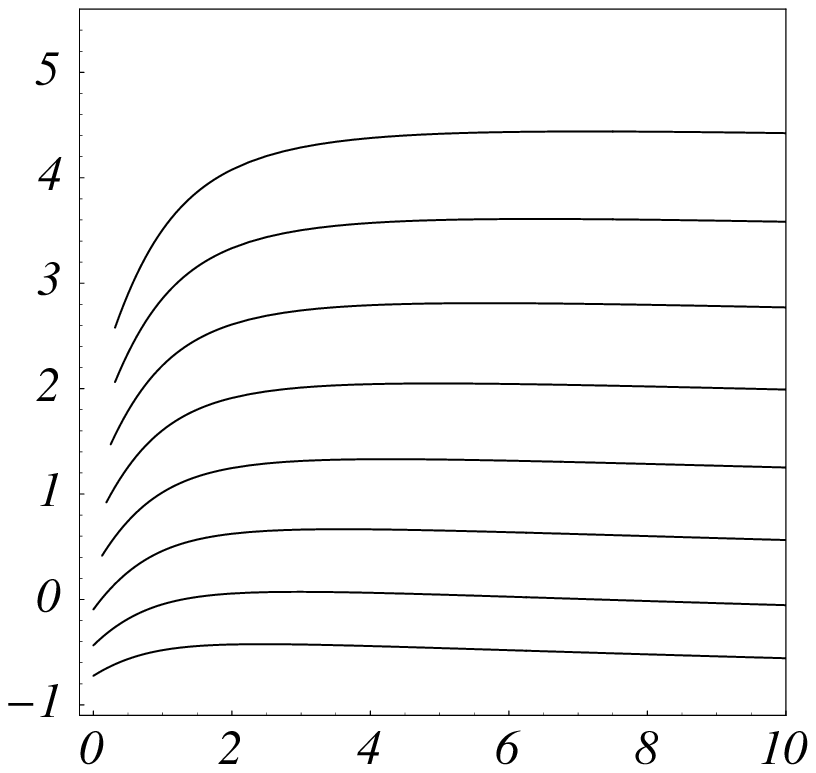,width=0.33\textwidth}}
\put(660,10){\epsfig{file=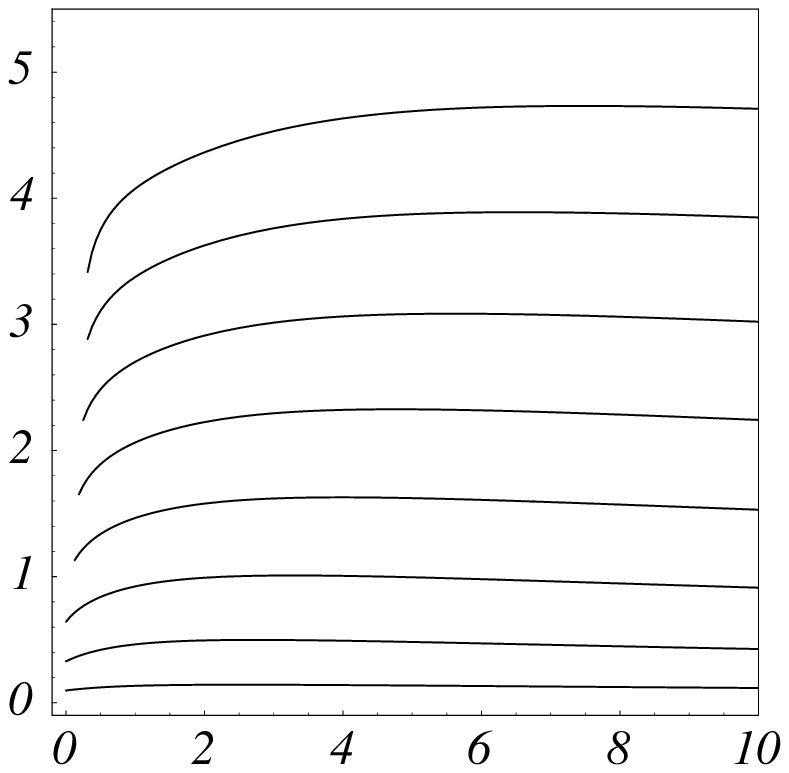,width=0.33\textwidth}}
\put(160,0){$\log_{10}b$}
\put(490,0){$\log_{10}b$}
\put(820,0){$\log_{10}b$}
\put(250,280){a) $\ \la_*$}
\put(100,220){$\lab$}
\put(380,280){b) $\ \log_{10} g_*$}
\put(720,280){c) $\ \tau_*$}
\end{picture}
\vspace{.5cm}
\caption{\label{pFP}
  Fixed points in $D=4+n$ dimensions with $n=0,...,7$
  (thin lines from bottom to top) 
  as a function of the cutoff parameter $b$ for the momentum cutoff 
  $r_{\rm mexp}(b)$ in Feynman gauge \cite{Reuter:1996cp}; 
  a) the cosmological constant $\lambda_*$ and $\lab$ (thick line);
  b) the gravitational coupling $g_*$;
  c) the dimensionless product $\tau_*=\lambda_*\, (g_*)^{2/(D-2)}$. }
\end{figure}
%*************************************************************************

\paragraph{Existence} 
A real, non-trivial, ultraviolet fixed point exists for all dimensions
considered.  Fig.~\ref{pFP} shows our results for $\la_*$ and $\log_{10}g_*$
based on a momentum cutoff in Feynman gauge \cite{Reuter:1996cp}
and scalar cutoff $r_{\rm mexp}(b)$, with parameter $b$ up to $10^{10}$.  For
small $b$, their numerical values depend strongly on $b$, while for large $b$,
they become independent thereof.  For $D>6$, the fixed point approaches the
boundary value $\la_*\to \lab$ for some $b_{\rm bound}>1$, beyond which the
momentum cutoff is no longer applicable.  Results similar to Fig.~\ref{pFP}
are found for all momentum cutoffs indicated above \cite{FischerLang}.

\paragraph{Continuity} 
The fixed points $\lambda_*$ and $g_*$, as a function of the dimension, 
are continuously connected with their perturbatively known counterparts 
in two dimensions \cite{Souma:1999at,Lauscher:2001ya,Litim:2003vp,Litim2005}. 

\paragraph{Uniqueness} 
This fixed point is unique in all dimensions considered.

\paragraph{Positivity of the gravitational coupling} 
The gravitational coupling constant only takes positive values at the fixed
point.  Positivity is required at least in the deep infrared, where gravity is
attractive and the renormalisation group running is dominated by classical
scaling.  Since the flow $\beta_g$ in \eqref{dglk} is proportional to $g$
itself, and stays finite for small $g$, it follows that renormalisation group
trajectories cannot cross the line $g=0$ for any finite scale $k$.  Therefore
the sign of $g$ is fixed along any trajectory, and positivity in the infrared
requires positivity already at an ultraviolet fixed point.  At the critical
dimension $D=2$, the gravitational fixed point is degenerate with the gaussian
one $(g_*,\lambda_*)=(0,0)$, and, consequently, takes negative values below
two dimensions.

\paragraph{Positivity of the cosmological constant} 
At vanishing $\lambda$, $\beta_\lambda$ is generically non-vanishing.
Moreover, it depends on the running gravitational coupling.  Along a
trajectory, therefore, the cosmological constant can change sign by running
through $\lambda=0$.  Then the sign of $\lambda_*$ at an ultraviolet fixed
point is not determined by its sign in the deep infrared.  We find that the
cosmological constant takes positive values at the fixed point, $\lambda_*>0$,
for all dimensions and cutoffs considered.  In pure gravity, the fixed point
$\lambda_*$ takes negative values only in the vicinity of two dimensions.
Once matter degrees of freedom are coupled to the theory, the sign of
$\lambda_*$ can change, $e.g.$ in four dimensions \cite{Percacci:2002ie}.  We
expect this pattern to persist also in the higher-dimensional case.

\paragraph{Dimensional analysis} In pure gravity (no cosmological constant
    term), only the sign of the gravitational coupling is well-defined, while
    its size can be rescaled to any value by a rescaling of the metric field
    $g_{\mu\nu}\to\ell\, g_{\mu\nu}$. In the presence of a cosmological
    constant, however, the relative strength of the Ricci invariant and the
    volume element can serve as a measure of the coupling strength. From
    dimensional analysis, we conclude that
\begin{equation}
\label{tau}
\tau_k =\bar\lambda_k \, (G_k)^{\frac{2}{D-2}}
\end{equation}
is dimensionless and invariant under rescalings of the metric field
\cite{Kawai:1992fz}.  Then the on-shell effective action is a function of
$\tau_k$ only. In the fixed point regime, $\tau_k$ reduces to
$\tau_*=\lambda_*\, (g_*)^{2/(D-2)}$.  In Fig.~\ref{pFP}c), we have displayed
$\tau_*$ using the cutoff $r_{\rm mexp}$ for arbitrary $b$. In comparison with
the fixed point values in Figs.~\ref{pFP}a,b), $\tau_*$ only varies very
mildly as a function of the cutoff parameter $b$, and significantly less than
both $g_*$ and $\lambda_*$. This shows that \eqref{tau} qualifies as a
universal variable in general dimensions. In Tab.~\ref{ABmeancomparison}, we
have collected our results for \eqref{tau} at the fixed point.  For all
dimensions shown, $\tau_*$ displays only a very mild dependence on the cutoff
function.

%*************************************************************************
% Figure 2 -- Universal exponents 
%*************************************************************************
\begin{figure}
\unitlength0.001\textwidth
\begin{picture}(1000,340)
\put(0,10){\epsfig{file=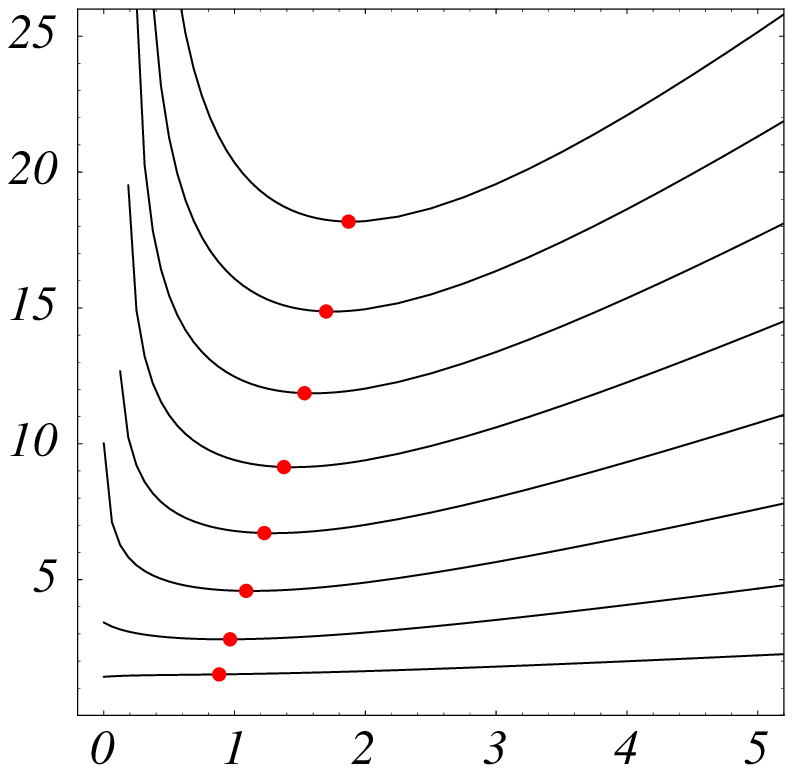,width=0.33\textwidth}}
\put(330,10){\epsfig{file=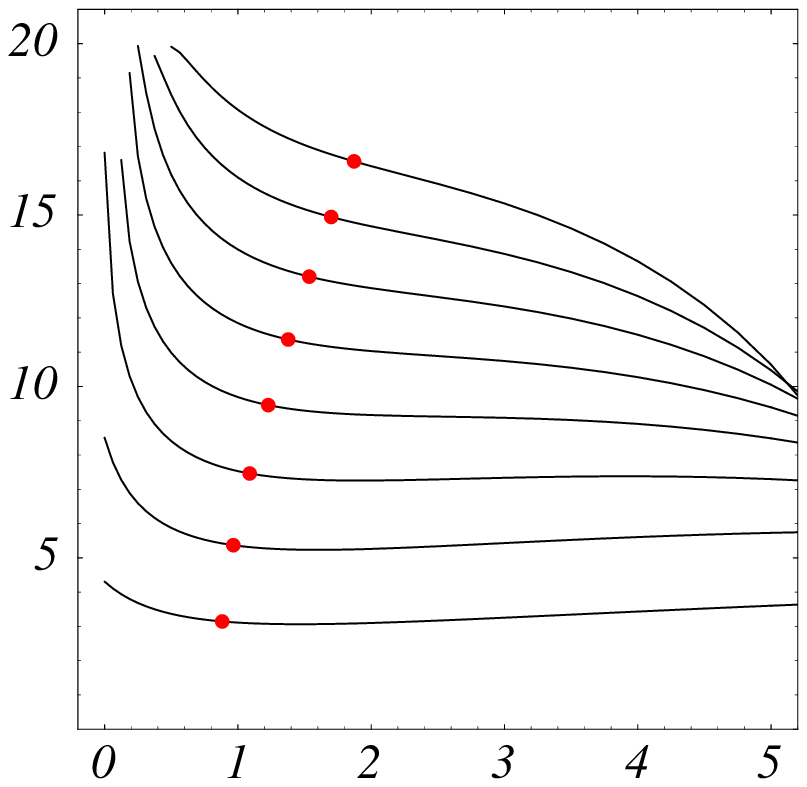,width=0.33\textwidth}}
\put(660,10){\epsfig{file=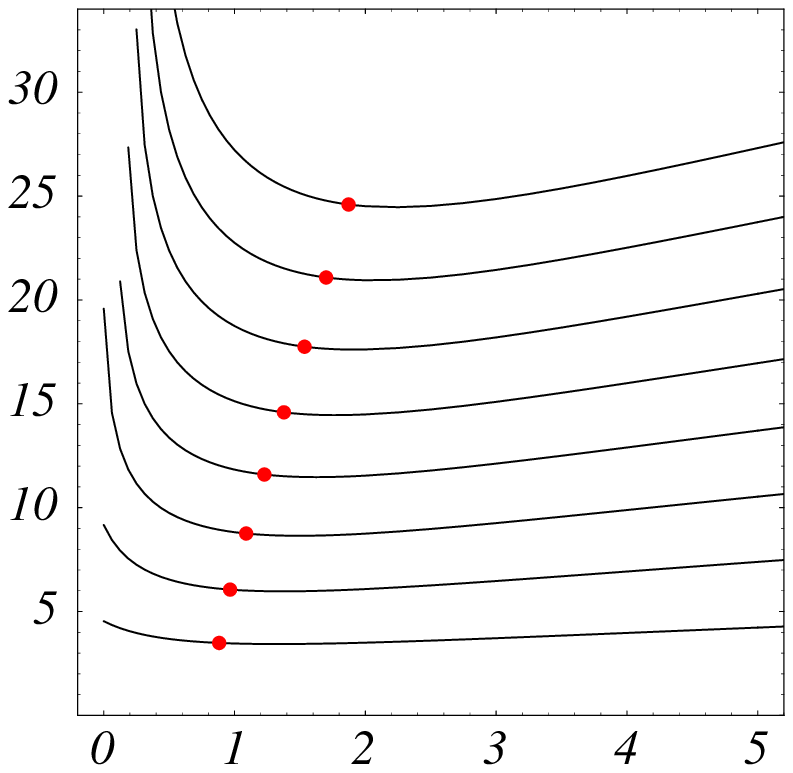,width=0.33\textwidth}}
\put(160,0){$\log_{10}b$}
\put(490,0){$\log_{10}b$}
\put(820,0){$\log_{10}b$}
\put(150,280){a) $\ \theta'$}
\put(580,280){b) $\ \theta''$}
\put(820,280){c) $\ |\theta|$}
\put(150,40){$n=0$}
\put(490,65){$n=0$}
\put(820,50){$n=0$}
\put(130,250){$n=7$}
\put(430,282){$n=7$}
\put(890,265){$n=7$}
\end{picture}
\caption{\label{AmexpThetas}
   Eigenvalues of the stability matrix at criticality in $D=4+n,\,n=0,\ldots7$
   (from bottom to top).
   The results obtaining from optimised cutoffs are indicated by the red dots.
   a) Real part $\theta'$; b) imaginary part $\theta''$; c) modulus $|\theta|$.}
\end{figure}
%*************************************************************************

\paragraph{Universality}
The fixed point values are non-universal. Universal characteristics of the
fixed point are given by the eigenvalues $-\theta$ of the Jacobi matrix with
elements $\partial_x \beta_y|_*$ and $x,y$ given by $\lambda$ or $g$,
evaluated at the fixed point. The Jacobi matrix is real though not symmetric
and admits real or complex conjugate eigenvalues.  For all cases considered,
we find complex eigenvalues $\theta\equiv\theta'\pm i\theta''$.  The real and
imaginary part and the modulus are displayed in Fig.~\ref{AmexpThetas} as
functions of the cutoff parameter $b$ for $r=r_{\rm mexp}$. The results for
the other cutoff functions are very similar.  In the Einstein-Hilbert
truncation, the fixed point displays two ultraviolet attractive directions,
reflected by $\theta'>0$.  Complex scaling exponents are due to competing
interactions in the scaling of the volume invariant $\int \sqrt{g}$ and the
Ricci invariant $\int \sqrt{g}R$ \cite{Litim2005}. The eigenvalues are real in
the vicinity of two dimensions, and in the large-$D$ limit, where the fixed
point scaling is dominated by the $\int \sqrt{g}$ invariant
\cite{Litim:2003vp}. $\theta'$ and $|\theta|$ are increasing functions of the
dimension, for all $D\ge 4$ \cite{Litim2005}.  For the dimensions shown here,
$\theta''$ equally increases with dimension.

 \paragraph{Convergence} The convergence of the results is assessed by
 comparing different orders in the expansion. The fixed point persists in the
 truncation where the cosmological constant is set to zero,
 $\lambda=\beta_\lambda=0$. Then, $\beta_g(g_*,\lambda=0)=0$ implies fixed
 points $g_*>0$ for all dimensions and cutoffs studied. The scaling exponent
 $\theta=-\partial\beta_g/\partial g|_*$ at $g_*$ is real and of the order of
 $|\theta|$ given in Tab.~\ref{ABmeancomparison}. The analysis can be extended
 beyond \eqref{EHk}, $e.g.$~including $\int \sqrt{g}R^2$ invariants and
 similar.  In the four-dimensional case, $R^2$ interactions lead to a mild
 modification of the fixed point and the scaling exponents
 \cite{Lauscher:2001ya,Lauscher:2002mb}.  It is conceivable that the
 underlying expansion is well-behaved also in higher dimensions.

%*************************************************************************
% Figure 3 -- Stability Parameter
%*************************************************************************
\begin{center}
\begin{figure}
\unitlength0.001\textwidth
\begin{picture}(670,340)
\put(0,10){\epsfig{file=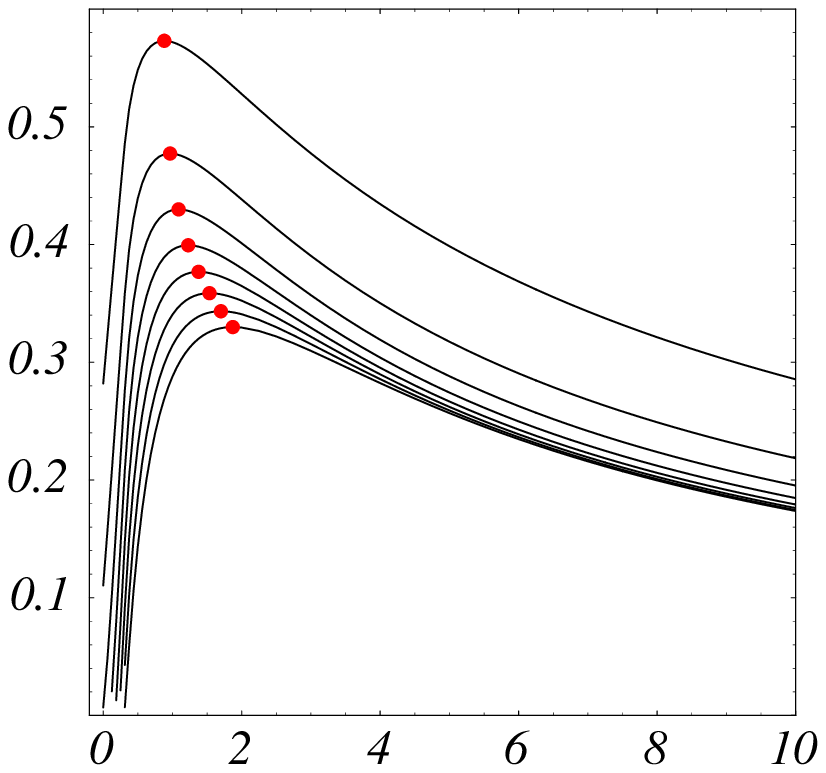,width=0.33\textwidth}}
\put(330,10){\epsfig{file=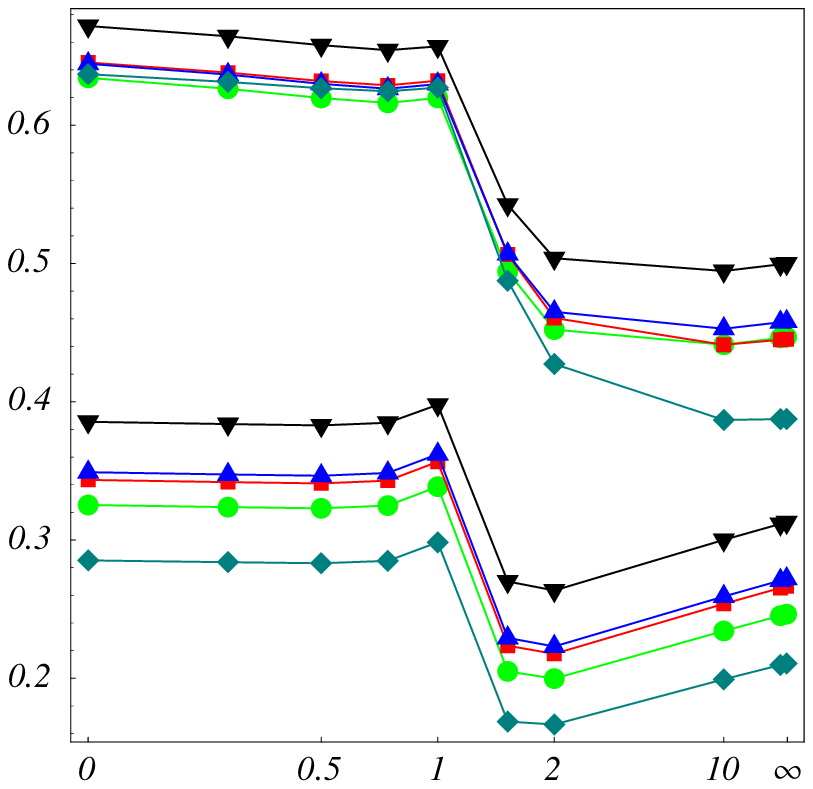,width=0.33\textwidth}}
\put(160,0){$\log_{10}b$}
\put(490,0){$\alpha$}
\put(110,280){a) }
\put(530,280){b) }
\end{picture}
\vspace{.5cm}
\caption{\label{pStability}
  Stability of renormalisation group flows. a) The parameter $1-\la_*/\lab$ as
  a function of the cutoff parameter $b$ for the momentum cutoff $r_{\rm
    mexp}$ in Feynman gauge \cite{Reuter:1996cp} (for more details, see
  \cite{Fischer:2006fz}); the red dot indicates the maximum. b) The maximum of
  $1-\la_*/\lab$ for different cutoff functions and gauge fixing parameters in
  a covariant background field gauge \cite{Lauscher:2001ya}. We note that
  $\alpha\le 1$ leads to more stable flows than $\alpha>1$.}
\end{figure}
\end{center}
%*************************************************************************

\paragraph{Stability}
The stability of the flow is controlled by the parameter $1-\la_*/\lab$,
displayed in Fig.~\ref{pStability} (see \cite{FischerLang}). As a function of
$b$, it displays a unique maximum for all classes of cutoffs studied.  In
Fig.~\ref{pStability} b), we have indicated the maximum values for
$1-\la_*/\lab$ as a function of the gauge fixing parameter. We note that the
stability of flows is larger for gauge fixing parameters $\alpha\le 1$. This
had to be expected, given that $\alpha=0$ is a non-perturbative fixed point of
the flow \cite{Litim:1998qi}.

%*************************************************************************
% Figure 4 -- Gauge and Cutoff Independence  
%*************************************************************************
\begin{center}
\definecolor{greygreen}{rgb}{0,.5,.5}
\begin{figure}
\unitlength0.001\textwidth
\begin{picture}(670,340)
\put(0,10){\epsfig{file=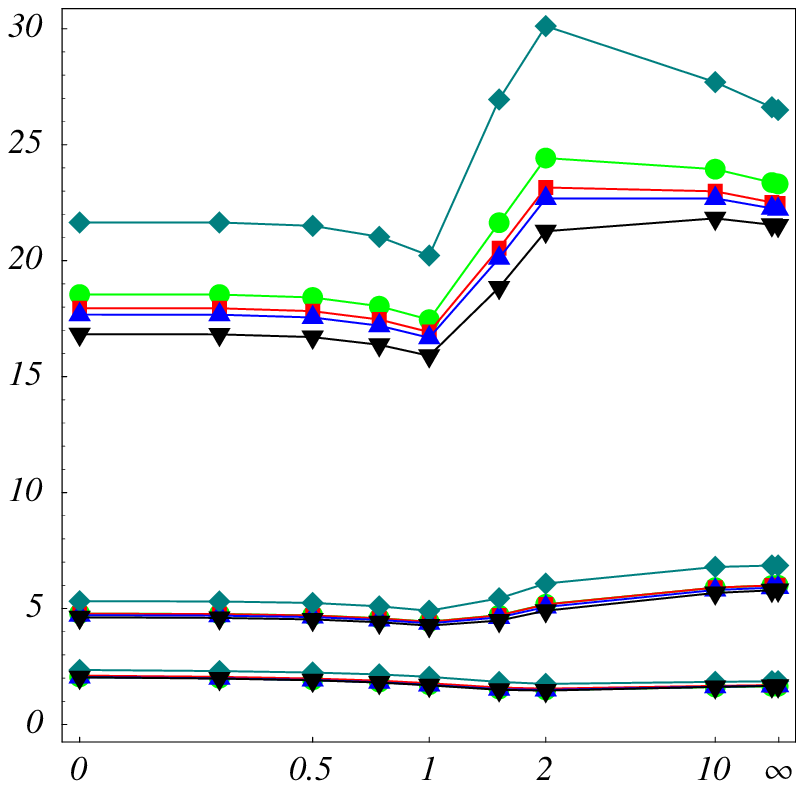,width=0.33\textwidth}}
\put(330,10){\epsfig{file=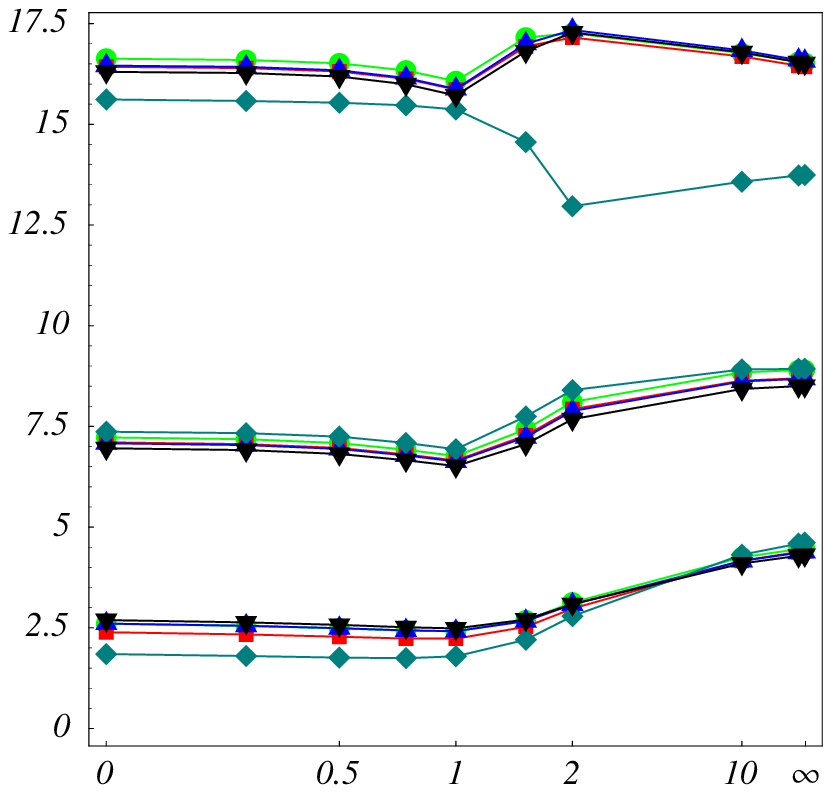,width=0.33\textwidth}}
\put(160,0){$\alpha$}
\put(490,0){$\alpha$}
\put(50,280){a) $\ \theta'$}
\put(390,250){b) $\ \theta''$}
\put(120,35){$n=0$}
\put(120,95){$n=2$}
\put(120,240){$n=7$}
\put(470,90){$n=0$}
\put(470,160){$n=2$}
\put(470,260){$n=7$}
\end{picture}
\caption{\label{GaugeIndep}
  Universal eigenvalues a) $\theta'$ and b) $\theta''$ for various gauge
  fixing parameter and cutoffs, with $n=0,2,7$ extra dimensions.  The five
  sets of data obtain from $r_{\rm mexp}$ \textcolor{green}{$\bullet$},
  $r_{\rm exp}$ \textcolor{red}{$\blacksquare$}, $r_{\rm
    mod}$\textcolor{blue}{$\blacktriangle$}, $r_{\rm pow}$
  \textcolor{greygreen}{$\blacklozenge$} and $r_{\rm opt}\,
  \blacktriangledown$.}
\end{figure} 
\end{center}
%*************************************************************************

\paragraph{Gauge-fixing independence} 
In the harmonic background field gauge \cite{Lauscher:2001ya}, the flow and
its solutions may depend spuriously on the gauge fixing parameter $\alpha$
\cite{FischerGauge}. We found a non-trivial fixed point and universal
eigenvalues for arbitrary gauge fixing parameter $\alpha \in [0,\infty)$, and
all classes of momentum cutoffs considered. Hence, the fixed point appears not
to be an artefact of the gauge fixing. Fig.~\ref{GaugeIndep} shows our
results for $\theta'$ and $\theta''$. For each $\alpha$, we first identified
the most stable flow parameters as in Fig.~\ref{pStability}, and then deduced
the corresponding values for $\theta'$ and $\theta''$. From
Fig.~\ref{pStability} b), we conclude that the stability of the flow is higher
for $\alpha\le 1$ as opposed to $\alpha>1$ \cite{Litim2005}. Furthermore, the
extremely weak $\alpha$-dependence of $1-\la_*/\lab$ in Fig.~\ref{pStability}
b), for $\alpha\le 1$, implies that physical observables should be independent
of $\alpha$ in this regime. This is indeed confirmed by our analysis in
Fig.~\ref{GaugeIndep}.

\paragraph{Cutoff independence} 
Non-trivial fixed points are found independently of the momentum cutoff,
$e.g.$~Fig.~\ref{pFP}, and independently of the gauge-fixing parameter
$e.g.$~Fig.~\ref{GaugeIndep}. The scaling exponents $\theta$, however, depend
spuriously on $R_k$ due to the truncation.  This dependence strictly vanishes
for the full, untruncated flow.  For best quantitative estimates of scaling
exponents we resort to an optimisation, following
\cite{Litim:2000ci,Litim:2002cf,FischerLang}, and use optimised values for
$b$, for each class of cutoffs given above.  Optimised flows have best
stability properties and lead to results closer to the physical theory
\cite{Litim:2002cf}. In Tab.~\ref{ABmeancomparison}, we show our results for
$\theta'$, $\theta''$ and $|\theta|$ obtained from the cutoffs $r_{\rm mexp},
r_{\rm exp},r_{\rm mod}$ and $r_{\rm opt}$ in Feynman gauge with $\alpha=1$,
and harmonic background field gauge with $0\leq \alpha \leq 1$. The remaining
spread in the numerical values indicates the remaining uncertainty due to the
truncation. On a quantitative level, it is found that $r_{\rm opt}$ with $b=1$
leads to flows with best stability. The variation in $\theta'$, $\theta''$ and
$|\theta|$ is reasonably small, and significantly smaller than the variation
with $b$ \cite{FischerLang}. With increasing $n$, the variation slightly
increases for $\theta'$ and $|\theta|$, and decreases for $\theta''$. The
expected error due to the truncation \eqref{EHk} is larger than the variation
in Tab.~\ref{ABmeancomparison}.  In this light, our results in the
Einstein-Hilbert truncation are cutoff independent.

%*************************************************************************
%Table
%*************************************************************************
  \begin{table}
\begin{tabular}{ccccccccc}
\hline
$n=D-4$  & 0 & 1 & 2 & 3 & 4 & 5 & 6 & 7 \\
\hline
$\theta'$ &1.48 -- 2.76&2.69 -- 3.11&4.26 -- 4.78&6.43 -- 6.89&8.19 -- 
                   9.34&10.5 -- 12.1&13.1 -- 15.2&15.9 -- 18.5\\
$\theta''$&2.23 -- 3.14&4.54 -- 5.16&6.52 -- 7.46&8.43 -- 9.46&10.3 -- 
                   11.4&12.1 -- 13.2&13.9 -- 15.0&15.7 -- 16.6\\
$|\theta|$&2.85 -- 3.49&5.31 -- 6.06&7.79 -- 8.76&10.4 -- 11.6&13.2 -- 
                   14.7&16.1 -- 17.9&19.1 -- 21.3&22.4 -- 24.9\\
\hline
\end{tabular}
\caption{Range of scaling exponents $\theta', \theta''$ and $|\theta|$,
  obtained from optimised
  momentum cutoffs 
      $r_{\rm mexp}, r_{\rm exp},r_{\rm mod}$ and $r_{\rm opt}$ in Feynman or
      harmonic background field gauge with $0\leq \alpha \leq 1$ (see text).}
\label{ABmeancomparison}
    \end{table}
%*************************************************************************

\paragraph{Anomalous dimension} 
The non-trivial fixed point implies a non-perturbatively large anomalous
dimension for the gravitational field, due to \eqref{dglk}, which takes
negative integer values $\eta=2-D$ at the fixed point.\footnote{Integer
  anomalous dimensions are known from other gauge theories at a fixed point
  away from their canonical dimension, $e.g.$~abelian Higgs
  \cite{Bergerhoff:1995zq} below or Yang-Mills \cite{Kazakov:2002jd} above
  four dimensions.}  The dressed graviton propagator ${\cal G}(p)$, neglecting
the tensorial structure, is obtained from evaluating $1/(Z_{N,k}\,p^2)$
for $k^2=p^2$. Therefore, the graviton propagator scales as
\begin{equation}
\label{Gscaling}
{\cal G}(p)\sim 1/p^{2(1-\eta/2)}\,,
\end{equation} 
which reads $\sim 1/(p^2)^{D/2}$ in the deep ultraviolet and should be
contrasted with the $1/p^2$ behaviour in the perturbative regime. The
anomalous scaling in the deep ultraviolet implements a substantial suppression
of the graviton propagator.  We verified the crossover behaviour of the
anomalous dimension from perturbative scaling in the infrared to ultraviolet
scaling by numerical integration of the flow \eqref{dglk}. 
More generally, higher
order vertex functions should equally display scaling characterised by
universal anomalous dimensions in the deep ultraviolet.  This is due to the
fact that a fixed point action $\Gamma_*$ is free of dimensionful parameters.

\paragraph{UV-IR connection} 
A non-trivial ultraviolet fixed point is physically feasible only if it is 
connected to the perturbative infrared domain by well-defined, finite 
renormalisation group trajectories.  
Elsewise, it would be impossible to connect the known low-energy physics 
of gravity with the putative high energy fixed point. 
A necessary condition is $\la_* < \lab$, which is fulfilled.  
Moreover, we have confirmed by numerical integration of the flow  
that the fixed points are connected to the perturbative infrared domain 
by well-defined trajectories in higher dimensions.

\paragraph{Running gravitational coupling}
The physically relevant trajectories connect the UV fixed point with the
vicinity of the gaussian fixed point in the infrared limit, and the running
gravitational coupling displays a cross-over between them.
In the deep ultraviolet, the non-trivial fixed point implies that the
gravitational coupling scales as
\begin{equation}\label{G-fix}
G_k=g_* / k^{D-2}\,.
\end{equation}
Hence, and in contrast to the regime of standard perturbation theory where
$G_k\approx$~const., the coupling becomes dimensionally suppressed at
a non-trivial ultraviolet fixed point ($k\to\infty$). Then gravity becomes
weakly coupled in the approach to the fixed point. This ``weakly coupled''
regime should be contrasted with asymptotic freedom of four-dimensional QCD,
where the weak coupling in the UV limit is due to a gaussian fixed point
rather than to the dimensional suppression \eqref{G-fix}.
 
%********************************************************************************
\subsection{Discussion} \label{disc}
%********************************************************************************

In summary, we have found a unique and non-trivial ultraviolet fixed point
with all the right properties for quantum gravity in more than four
dimensions. Within the present truncation, the fixed point exists
independently of the cutoff and the gauge fixing. Furthermore, for
sufficiently stable flows, we have established that universal scaling
exponents are even quantitatively independent of the cutoff and the gauge
fixing parameter. This constitutes a highly non-trivial consistency check for
the result. Together with the renormalisation group trajectories which connect
the fixed point with the perturbative infrared domain, our findings provide a
viable realisation of Weinberg's asymptotic safety scenario in terms of the
metric degrees of freedom.  \\[-1ex]

In future work, these studies should be extended to include higher powers in
the Riemann tensor, the Ricci scalar, derivatives thereof, and matter degrees
of freedom. In the four-dimensional case, $R^2$ interactions lead only to
minor modifications of the fixed point and its scaling properties
\cite{Lauscher:2002mb}. It is important to confirm this pattern in higher
dimensions, where $R^2$ interactions are relevant already in perturbation
theory.  If the fixed point persists in extensions, quantum gravity can well
be formulated as a fundamental theory in the metric degrees of freedom.
Furthermore, it would be useful to have an independent verification of the
fixed point based on $e.g.$~lattice simulations. Recent progress in the
four-dimensional case seems to suggest that Monte-Carlo studies of
higher-dimensional Einstein gravity within Regge calculus or causal dynamical
triangulation is in reach \cite{Hamber:1999nu,Ambjorn:2004qm}.
\\[-1ex]

Finally, we highlight main implications of the above fixed point for
phenomenological particle physics models with compact extra dimensions and low
scale quantum gravity. If realised in Nature, quantum gravity at the TeV scale
could be accessible experimentally in hadron colliders. In these models,
standard model particles live on a four-dimensional brane whereas gravity
propagates in a $D=4+n$ dimensional bulk.  Under the assumption that standard
model particles do not spoil the fixed point, we can neglect their presence
for the following considerations.  Without loss of generality, we consider $n$
extra spatial dimensions with compactification radius $L$.  The
four-dimensional Planck scale $M_{\rm Pl}$ is related to the $D$-dimensional
(fundamental) Planck mass $M_D$ and the radial length $L$ by the relation
$M^2_{\rm Pl}\sim M^2_{D} (M_D\,L)^{n}$, where $L^n$ is a measure for the
extra-dimensional volume.  A low fundamental Planck scale $M_D\ll M_{\rm Pl}$
therefore requires the scale separation $1/L\ll M_D$ which states that the
radius for the extra dimensions has to be much larger than the fundamental
Planck length $1/M_D$. For momentum scales $k\ll 1/L$, where $\eta\approx 0$,
the hierarchy implies that the running couplings scale according to their
four-dimensional canonical dimensions. At $k\approx 1/L$, the size of the
extra dimensions is resolved and, with increasing $k$, the couplings display a
dimensional crossover from four-dimensional to $D$-dimensional scaling. Still,
the graviton anomalous dimension stays small and gravitational interactions
$G_k\approx$~const.~are perturbatively weak.  This dimensional crossover is
insensitive to the fixed point in the deep ultraviolet.  In the vicinity of
$k\approx M_D$, however, the graviton anomalous dimension displays a
classical-to-quantum crossover from the gaussian fixed point $\eta\approx 0$
to non-perturbative scaling in the ultraviolet $\eta\approx 2-D$. This
crossover takes place in the full $D$-dimensional theory.  In the transition
regime, the propagation of gravitons is increasingly suppressed.  In addition,
the gravitational coupling becomes weak, following \eqref{G-fix}. Therefore,
the onset of the fixed point scaling cuts off gravity-mediated processes with
characteristic momenta at and above $M_D$, and provides dynamically for an
effective momentum cutoff of the order of $M_D$.  This dynamical suppression,
and a decreasing anomalous dimension $\eta<0$ in the transition regime, can be
seen as signatures for the non-trivial fixed point.  We conclude that the
gravitational fixed point could be detectable in experimental setups sensitive
to the TeV energy range, $e.g.$~in hadron colliders, provided that the
fundamental scale of gravity is as low as the electroweak scale.

\paragraph{Acknowledgements}
The work of DFL is supported by an EPSRC Advanced Fellowship.

\end{document}